\documentclass[iop]{emulateapj}

\usepackage{graphicx}
\usepackage{hyperref}
\usepackage{color}

\newcommand{\logg}{$\log{g}$}
\newcommand{\teff}{$T_{\rm eff}$}
\newcommand{\msun}{$M_\odot$}

\newcommand{\rsun}{$R_\odot$}

\newcommand{\numax}{\mbox{$\nu_{\rm max}$}}
\newcommand{\numaxsun}{\mbox{$\nu_{\rm max,\odot}$}}
\newcommand{\Dnu}{\mbox{$\Delta \nu$}}
\newcommand{\Dnusun}{\mbox{$\Delta \nu_{\odot}$}}

\shorttitle{Masses of Retired A Stars}
\shortauthors{Ghezzi \& Johnson}

\begin{document}

\title{Beyond the Main Sequence: Testing the accuracy of stellar masses \\ predicted by the PARSEC evolutionary tracks}

\author{
Luan Ghezzi\altaffilmark{1}
\& John Asher Johnson\altaffilmark{1}
}

\altaffiltext{1}{Harvard-Smithsonian Center for Astrophysics, 60 Garden Street, Cambridge, MA 02138 USA; lghezzi@cfa.harvard.edu}

\begin{abstract}
Characterizing the physical properties of exoplanets, and understanding their formation and orbital evolution requires precise and accurate knowledge of their host stars. Accurately measuring stellar masses is particularly
important because they likely influence planet occurrence and the architectures of planetary systems. Single main-sequence stars typically have masses estimated from evolutionary tracks, which generally provide accurate results due to their extensive empirical calibration. However, the
validity of this method for subgiants and giants has been called into question by recent studies, 
with suggestions that the masses of these evolved stars could have been overestimated.
We investigate these concerns using a sample of 59 benchmark evolved stars with model-independent masses (from binary systems or asteroseismology) obtained from the literature. We find very good agreement between these benchmark
masses and the ones estimated using evolutionary tracks. The average fractional difference in the mass interval $\sim$0.7 -- 4.5 \msun\, is consistent with zero (-1.30 $\pm$ 2.42\%), with no significant trends in the residuals relative to the input parameters. 
A good agreement between model-dependent and -independent radii (-4.81 $\pm$ 1.32\%) and surface gravities (0.71 $\pm$ 0.51\%) is also found. The consistency between independently determined ages for members of binary systems adds further support for the accuracy of the method employed to derive the stellar masses.
Taken together, our results indicate that determination of masses of evolved stars using grids of evolutionary tracks 
is not significantly affected by systematic errors, and is thus valid for estimating the masses of isolated stars beyond the main sequence.
\end{abstract}

%% Keywords should appear after the \end{abstract} command. The uncommented
%% example has been keyed in ApJ style. See the instructions to 


\section{Introduction}
\label{introduction}

A complete understanding of the formation and evolution of planetary systems 
requires a detailed knowledge of the physical properties of the planets and their 
host stars. 
A subject of intense study within exoplanetary science are the nature of the links between planets and the stars they orbit. Planets form from the same molecular cloud material from which their host stars form, so planets can be thought of as the leftover relics of the star-formation epoch. Similarly, stars can be studied as relics of the environment from which planets 
formed long ago \citep{lba04,fv05,kk08,johnson08}. 
Relationships between planetary systems observed today and their host star's physical properties therefore provide important constraints on theories of planet formation and orbital evolution.

The two key physical properties that govern stellar evolution are mass and chemical composition, and these characteristics of host stars are thought to be linked to the formation and orbital evolution that produced the architectures of planetary systems seen today. Many independent studies have shown that the occurrence of 
giant planets around FGK dwarfs and post-main-sequence stars such as subgiants strongly increases with the stellar 
iron abundance (e.g., \citealt{gonzalez97,santos04,fv05,johnson10a}). However, 
there exists much debate as to whether this correlation between metallicity and giant planet occurrence exists for stars on the red giant branch (e.g,
\citealt{johnson10a, maldonado13,mortier13,jofre14, reffert15}) or stars that host planets with masses less than that of Neptune (e.g., 
\citealt{ghezzi10b,sousa11,buchhave14}). Evidence for possible connections between 
the presence of planets and other peculiarities in the chemical abundances of 
their stellar hosts are still ambiguous (see, e.g., 
\citealt{adibekyan12,adibekyan14,figueira14,teske14}; and references therein).

Also still unclear is the role of stellar mass on the formation and evolution of
extrasolar planets. Planet-search programs based on the radial velocities (RVs)
technique have mainly targeted Solar-type stars because they are relatively numerous
and bright in the Solar neighborhood, and their spectra have a considerable number of unblended, narrow metallic
lines, allowing more precise measurements of both the Doppler shifts and stellar
properties \citep[e.g.][]{galland05,lagrange09,becker15}. One direct consequence of this selection criterion is a relatively small
range of stellar masses included among the targets of the largest existing planet-search programs ($\sim$0.7 --1.4~\msun). The extension of this
interval to lower-mass M dwarfs was a natural extension of the exoplanet searches
(e.g., \citealt{johnson07a,bonfils13}), given these stars are more numerous 
in the Galaxy and allow robust detections of Earth-like planets with the current
instrumental capabilities (e.g., \citealt{mayor14,quintana14}). Nevertheless, the
monitoring and characterization of M dwarfs has been challenging because they are 
faint and their optical spectra are completely dominated by molecular bands \citep[e.g.][]{bean06, maness07}. 
Dedicated ongoing and future surveys will
progressively overcome these issues (e.g., \citealt{berta13,quirrenbach14}).

A completely different difficulty is encountered for planet searches around massive stars on the main sequence
($\gtrsim$1.5 \msun). The hotter effective temperatures and higher rotational
velocities of early F and A stars produce a spectrum with fewer and much broader
lines, hindering the measurement of precise radial velocities and the reliable
detection of extrasolar planets \citep{galland05, becker15}. However, as these massive stars evolve off the main sequence 
towards the red giant branch (RGB), they cool and slow down (\citealt{doNascimento00}).
As a result, their spectra then show many narrow lines which allow the measurement of Doppler
shifts caused by extrasolar planets. Therefore, analyzing the evolved counterparts
of main-sequence F- and A-type dwarfs is currently the best alternative to investigate planet
occurrence around more massive stars\footnote{Direct imaging is a promising technique
to detect giant planets around B-A stars, but so far only a handful of them were 
found (see, e.g., \citealt{cj11,nielsen13} and references therein)}.

To this end, more than 1000 evolved stars have been monitored by several surveys
(see \citealt{niedzielski15} and references therein), resulting in the detection of
more than 100 planets around them (\citealt{jofre14}). Although this sample is
relatively small when compared to the entire sample of discovered planets, 
some interesting properties already started to emerge, such as a
paucity of planets at short orbital distances and large eccentricities
(\citealt{sato07,bowler10,jones14}) and the possible lack of a giant planet -- metallicity correlation
(\citealt{ghezzi10a,maldonado13,jofre14}; but see also \citealt{reffert15}). 

Of particular interest to this work is 
the higher occurrence rate of Jovian planets around more massive stars measured by
\cite{johnson10a}. The analysis of 1194 stars with masses in the range 0.2 -- 2.0 \msun\, 
not only provided additional confirmation to the well-established giant planet-metallicity 
correlation, but also revealed that the occurrence rate of these planets
increases linearly from $\sim$3\% for M dwarfs to $\sim$14\% for A stars. 
This is a very important result because it adds new constraints to the planet
formation theories and also provides guidance to ongoing and future surveys 
monitoring evolved stars (radial velocities and transits) or A dwarfs (direct 
imaging -- \citealt{cj11}). 

The determination of masses for these isolated evolved stars relied on the
comparison of observed properties with stellar evolutionary tracks 
(\citealt{johnson10a,johnson11,johnson13}).
The different models usually adopted in this method were subjected to numerous improvements 
and validation tests over the last few decades, leading to more precise predictions of the stellar 
parameters and also a better understanding of the processes involved in the evolution of stars
(e.g., \citealt{vandenberg85,andersen88,a91,andersen91,pols97,demarque04,torres10,bressan12,brogaard12,garcia14,torres15}).    
For field stars on the main sequence, this technique for determining masses generally provides accurate 
results (e.g., \citealt{torres10}) that are also fairly consistent with 
other independent estimates (\citealt{pinheiro14}; but note the possible issues with spectroscopic 
masses for stars with M $\gtrsim$ 1.2 \msun).  

The reliability of the application of this method to subgiants and giants is,
however, more uncertain. Although stellar evolution models were able to successfully describe 
some binary systems with at least one evolved component (e.g., \citealt{andersen88,andersen91,torres15}),
there were suggestions that the masses of planet-hosting subgiants and giants could have been overestimated 
(by up to 50\%) due to systematic errors on their atmospheric parameters or the models themselves 
(\citealt{lloyd11,lloyd13,sw13}). 
Additional evidence of possibly overestimated masses for evolved stars with planets (by up to 100\%) were recently 
presented by \cite{sousa15} (but note that a possible explanation for some of the 
problematic stars is given in Section 4.2 of \citealt{tt15}).
In spite of the growing number of evolved stars 
with precisely determined masses (through orbital solutions in binary systems or
asteroseismology of field or cluster stars), the evolutionary tracks continue to be
main method to determine this fundamental parameter for single subgiants and giants.
Thus, possible errors in the results would have implications on many different studies,
ranging from the formation and architectures of planetary systems to Galactic chemical
and dynamical evolution, and have to be carefully investigated.

We and others have been conducting a study to precisely constrain the masses of evolved
stars using different input parameters and techniques. The first results of this
effort for the bright nearby subgiant star HD 185351 were presented by \cite{johnson14}. In 
this work, we check the accuracy of the evolutionary tracks method using a sample 
of benchmark subgiants and giants with accurate masses. This sample and the 
literature masses for their masses are described in Section \ref{sample}, 
while the determination of corresponding stellar masses from evolutionary tracks is
discussed in Section \ref{tracks}. In Section \ref{discussion}, we compare the
two sets of masses and show that the evolutionary tracks method do not
seem to have any systematic errors that would overestimate the masses of evolved
stars. Finally, our concluding remarks are presented in Section \ref{conclusion}.

\section{Sample of Benchmark Stars}
\label{sample}

Our sample of benchmark stars consists of 59
subgiants and giants with precise masses determined with methods that are based on
minimal assumptions and physical modeling. In this study, we focus on dynamical
masses measured for stars in binary systems and asteroseismic masses calculated
with scaling relations and the \textit{direct method} (i.e., without the usage of
[Fe/H] and grids of models to further constrain the evolutionary parameters; see, 
e.g., \citealt{gai11}). 

\subsection{Stars in Binary Systems}
\label{binaries}

We performed an extensive literature search for detached binary systems 
in which at least one of the components is an evolved star. Most of them were found
with the help of DEBCat\footnote{\url{http://www.astro.keele.ac.uk/jkt/debcat/}}
\citep{southworth14}. Only subgiants and giants that have available values for
effective temperature (\teff), metallicity ([Fe/H]), $V$ magnitude and
parallax ($\pi$) or distance ($d$) were selected. These are the input parameters
necessary to obtain masses with the evolutionary track method (see Section
\ref{tracks}). We removed any stars with metallicities determined through the
best fit to evolutionary tracks or isochrones (e.g., \citealt{lacy12,ratajczak13}) 
in order to avoid dual dependencies on the model tracks, e.g. using [Fe/H] derived 
from one evolution model grid to interpolate onto another model grid.

\begin{figure}
%\epsscale{0.9}
%\plotone{f1.eps}
\includegraphics[trim={0.2cm 0.8cm 1.5cm 0.4cm}, scale=0.8]{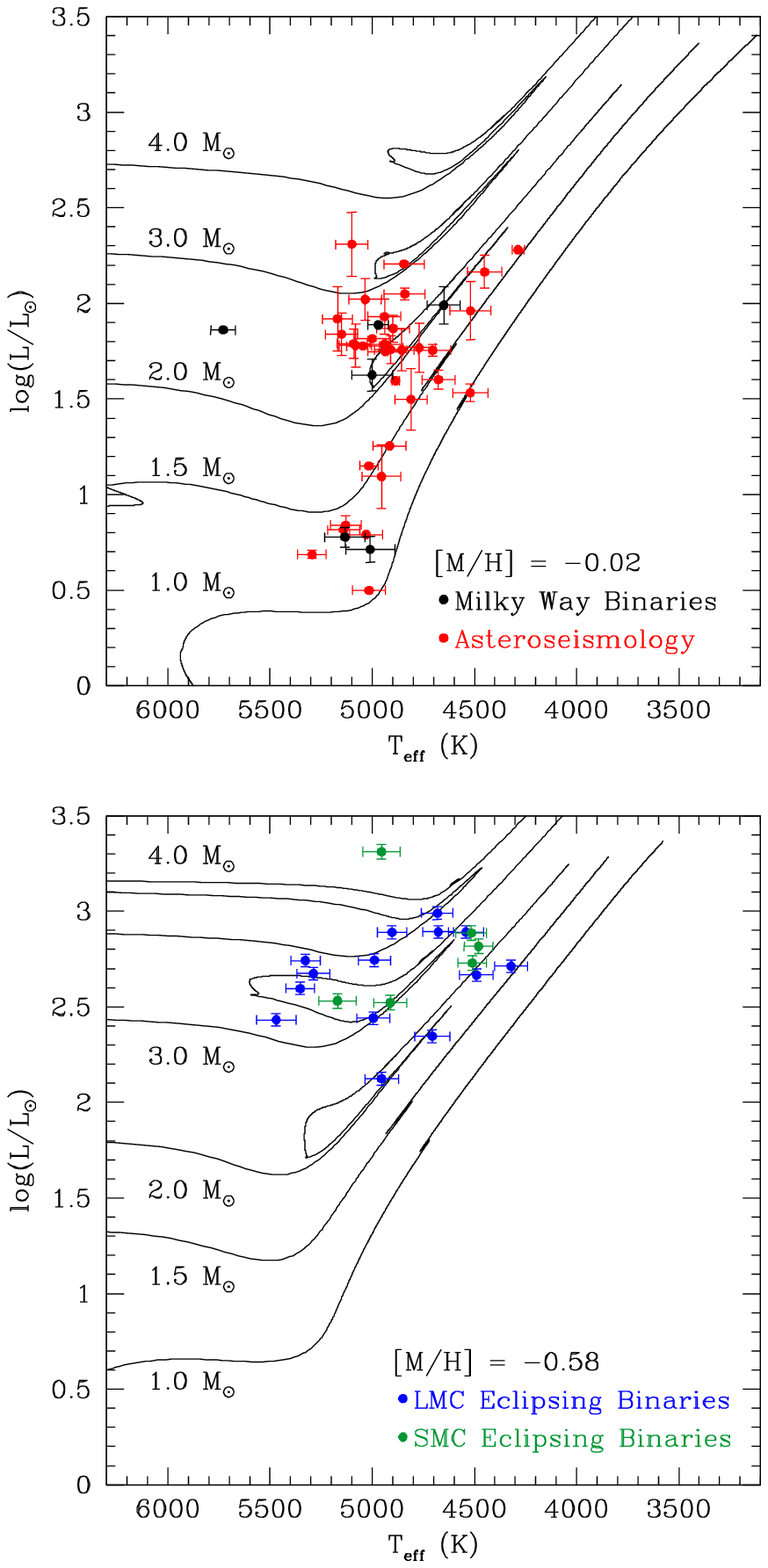}
\caption{HR diagrams showing the positions of stars in our sample. The upper
panel shows the MW binaries (black points) and field stars with asteroseismology (red
points). The lower panel presents the eclipsing binaries (EBs) from LMC (blue
points) and SMC (green points). In all panels, the solid lines
represent PARSEC evolutionary tracks \citep{bressan12} for 1.0 \msun, 1.5 \msun, 2.0 \msun, 3.0 \msun\, and 4.0 \msun,
and representative metallicities for the depicted samples.}
\label{hr_diagram}
\end{figure}

Our selection criteria resulted in a sample of 26 evolved stars in 16
binary systems in the Milky Way (MW), Small Magellanic Cloud (SMC) and Large
Magellanic Cloud (LMC). These stars are shown in Tables \ref{table_sample} and \ref{table_results_param} 
along with their fundamental parameters, respective uncertainties and literature references.
In some cases, no uncertainty was available for the $V$ magnitudes and we adopted $\sigma(V)$ = 0.01
for MW stars because this is the typical accuracy of the \textit{Hipparcos} measurements \cite{perryman97}.
For LMC stars, we adopted a conservative uncertainty $\sigma(V)$ = 0.02 because the accuracy of photometric 
calibrations in the OGLE project is better than 0.02 mag \citep{udalski08}. 
We also assumed $E(B-V)$ = 0.000 for all stars closer than $\sim$60 pc. 
Their positions in a HR diagram can be seen in Figure \ref{hr_diagram}. 
Dynamical stellar masses and radii were directly determined in the original source 
papers with precisions of $\lesssim$3\% and $\lesssim$7\%, respectively, through the 
analysis of radial velocities (RVs) and light curves. 
The masses range from $\sim$1.2 to $\sim$4.5 \msun.

\subsection{Stars with Asteroseismic Parameters}
\label{seism}

We conducted a separate literature search for evolved single stars with
measurements of the two global asteroseismic parameters: the frequency of maximum
power \numax\, and the large frequency separation \Dnu. Once again, we kept only those
stars which also had reported, evolution-model-independent values for \teff, [Fe/H], $V$ magnitude and parallax
($\pi$) or distance ($d$) (see Section \ref{binaries}). The last requirement,
in particular, dramatically reduces the number of stars suitable for our study. 

We calculated asteroseismic masses and radii with the \textit{direct method}, using published values of \numax, 
\Dnu\, and \teff\, as input to the scaling relations (e.g., \citealt{kb95}) written as:

\begin{equation}
\label{seism_mass}
\frac{M}{M_{\odot}} \simeq 
\Bigg(\frac{\nu_{\rm max}}{\nu_{\rm max,\odot}}\Bigg)^{3} \Bigg(\frac{\Delta\nu}{\Delta\nu_{\odot}}\Bigg)^{-4}
\Bigg(\frac{T_{\rm eff}}{T_{\rm eff,\odot}}\Bigg)^{3/2}
\end{equation}

\begin{equation}
\label{seism_radius}
\frac{R}{R_{\odot}} \simeq 
\Bigg(\frac{\nu_{\rm max}}{\nu_{\rm max,\odot}}\Bigg) \Bigg(\frac{\Delta\nu}{\Delta\nu_{\odot}}\Bigg)^{-2}
\Bigg(\frac{T_{\rm eff}}{T_{\rm eff,\odot}}\Bigg)^{1/2}.
\end{equation}

\noindent The reference solar asteroseismic parameters \numaxsun\, = 3090 $\pm$ 30 $\mu$Hz 
and \Dnusun\, = 135.1 $\pm$ 0.1 $\mu$Hz \citep{huber11} were adopted in the above
equations, as well as the canonical value $T_{\rm eff,\odot}$ = 5777 K. We note that,
while these results are independent from stellar evolution models, they do rely
on scaling of solar p-modes to stars of different masses, metallicities and evolutionary states.
Recent studies have shown that the usage of the scaling relations for metal-poor or evolved
stars require some caution (e.g., \citealt{miglio12,epstein14}).
However, many different tests indicate that these scaling relations provide masses
and radii with accuracies $\sim$10\% and $\sim$5\%, respectively (see discussion in
Section 2.4.1 of \citealt{johnson14}). No corrections were applied to the scaling
relations because there is still no consensus which of the proposed ones works 
best (e.g., \citealt{brogaard14}). Nevertheless, the shifts in $M$ and $R$ that would 
be caused by adopting the available corrections are discussed in Section \ref{discussion}.

We should also note that Equations \ref{seism_mass} and \ref{seism_radius} contain 
the stellar effective temperature, which in turn can be determined using model-dependent
techniques (for instance, spectroscopic \teff\, depends on model atmospheres). However, we
note that the dependence of the stellar radius on \teff\, in Equation~\ref{seism_radius} is weak relative to the other terms. Moreover, 
the uncertainties on the values of masses and radii caused by typical errors on the
effective temperatures ($\sim$80 K or $\sim$2\% considering a typical \teff = $\sim$4900~K) are negligible (of order 1\%).

As one of our goals is to define a statistically significant sample of benchmark stars, 
we decided to remove stars for which the uncertainties in the asteroseismic
masses were higher than 20\%. After this cut, we obtained a subsample of 33 evolved 
stars with asteroseismic parameters and masses with typical precisions ranging 
from 5\% to 20\%. These relatively large errors could be decreased if we used
a grid-based asteroseismic method \citep{gai11}, in which the stellar properties predicted by a grid of stellar evolution models are converted to the asteroseismic parameters $\nu_{\rm max}$ and $\Delta \nu$, and then compared to the observed values. But the goal of this work is to avoid any model dependencies in the reference parameters of the benchmark stars sample.

Our asteroseismic benchmark stars are shown in Tables \ref{table_sample} and \ref{table_results_param}.
As for the binaries, $\sigma(V)$ = 0.01 was adopted for stars with no uncertainties available for $V$.
We assumed $E(B-V)$ = 0.000 for all stars closer than $\sim$60 pc as well as 10 eight more distant
stars for which no values were available in the literature.
When no uncertainties were not reported for the asteroseismic parameters, arbitrary errors of 5\% and 2\% were adopted for \numax\, and \Dnu, respectively (following \citealt{bruntt10}). 
The positions of the stars in a HR diagram is shown in Figure \ref{hr_diagram} and their masses range from roughly 0.7~\msun\ 
to 4.0~\msun. One star (KIC\,8410637) is common to both the binary and asteroseismic samples. 
In this case, we adopt the more precise dynamical stellar mass and radius\footnote{The dynamical and asteroseismic masses,
radii and surface gravities agree within 1.6$\sigma$ (see Table \ref{table_results_param}).}.

\subsection{Final Sample}
\label{final_sample}

Our final sample consists of 59 benchmark stars with precise dynamical or asteroseismic
masses. According to model-dependent classifications provided in the reference papers or to 
Figure \ref{hr_diagram}, we can see that most of our stars are on the red giant branch (RGB) or 
on the red clump (RC), which are also the regions on the H--R diagram targeted by many planet-search 
surveys that focus on evolved stars (e.g., \citealt{hatzes05,sato05,lovis07,lee11,omiya12,niedzielski15,reffert15}).
All of them have the required parameters to allow an independent determination 
of their masses using stellar evolution models (see section \ref{tracks}).

As explained in the previous sections, some stars were removed from our sample because
they did not fulfill one or more of our selection criteria. For completeness, those 
stars are listed in Table \ref{table_removed_stars} along with the selected references
and the reasons for their exclusion. It is also 
worth noting that we may have overlooked a few stars despite our best efforts to conduct a thorough literature search\footnote{Please, contact the authors if you have identified such
a case.}. However, we consider that our current sample is statistically
robust, and the addition of a few objects will not likely affect our conclusions. 

\section{Stellar Masses from Evolutionary Tracks}
\label{tracks}

We derived model-dependent stellar masses for all 59 of our benchmark stars using stellar evolution model tracks. In this method, measured stellar properties (typically, \teff,
log--luminosity $\log_{10}{(L/L_{\odot})}$, and metallicity [Fe/H]) are compared with the 
properties predicted by evolution models calculated for different initial masses 
and chemical compositions \citep[e.g.][]{padova, dartmouth, yy}. A probabilistic analysis then provides the best match between
the observed and theoretical properties of the star, allowing the determination of 
its mass (as well as radius and age). This method is known to work well for solar-type
stars on the main sequence (e.g., \citealt{torres10}), but has not been extensively
tested on for stars in advanced evolutionary states \citep[however, see][]{johnson13, johnson14}. Testing these model grids using a large sample of evolved stars is the primary goal of this study.

We adopted the implementation of the method contained in the PARAM code, which was
kindly provided by Leo Girardi\footnote{For the web interface maintained by Leo 
Girardi at the Osservatorio Astronomico di Padova, visit
http://stev.oapd.inaf.it/cgi-bin/param.}. Estimates of the mass, radius, \logg\, and age
are obtained through a Bayesian estimation method using different
priors, input parameters and isochrones \citep{dasilva06}. We adopted the default options for the Bayesian
priors and selected the new PARSEC isochrones from \cite{bressan12}. We also chose the
analysis of stars with known parallaxes, for which we provided the additional input
parameters \teff, [Fe/H] and V magnitude (see Table \ref{table_sample}). Extinction corrections $A_{V}$
were applied to the V magnitudes of all stars farther than 60 pc for which a reddening value 
$E(B-V)$ was available in the literature. The standard relation $A_{V} = 3.1E(B-V)$
was adopted for the conversion of reddening to $V$-band extinction.

It is worth noting that the Padova models (in their most recent version, PARSEC; \citealt{bressan12})
were chosen because its wide usage in the literature, including the estimation the properties of evolved stars (e.g., 
\citealt{johnson06,johnson07b,dollinger07,dollinger09,demedeiros09,ghezzi10a,ghezzi10b}).
Tests with other grids of evolutionary tracks are beyond the scope of the present work,
but will be adressed in a future contribution using the same sample of benchmark stars.

The masses derived with the evolutionary tracks are shown in Table 
\ref{table_results_param} along with the other output parameters of PARAM and their
respective uncertainties. For the giant star $\nu$\,Ind, the error on [Fe/H] was arbitrarily increased 
to 0.20 dex because the PARAM code was not returning valid solutions with the original value ($\sigma_{\rm [Fe/H]} = 0.07$). 
After some tests, it was clear that PARAM was having issues due to the relatively
low metallicity and small original uncertainty. The precision of the masses
obtained with the evolutionary tracks method varies from $\sim$2\% to $\sim$29\%, with a
typical (median) value of $\sim$10\%. The relative uncertainties on the masses are similar for
the samples of binaries and asteroseismic targets and increase for stars with lower effective
temperatures and larger errors on their metallicities. 

\section{Results and Discussion}
\label{discussion}

\subsection{Masses}
\label{discussion_masses}

\begin{figure*}
\epsscale{0.9}
\plotone{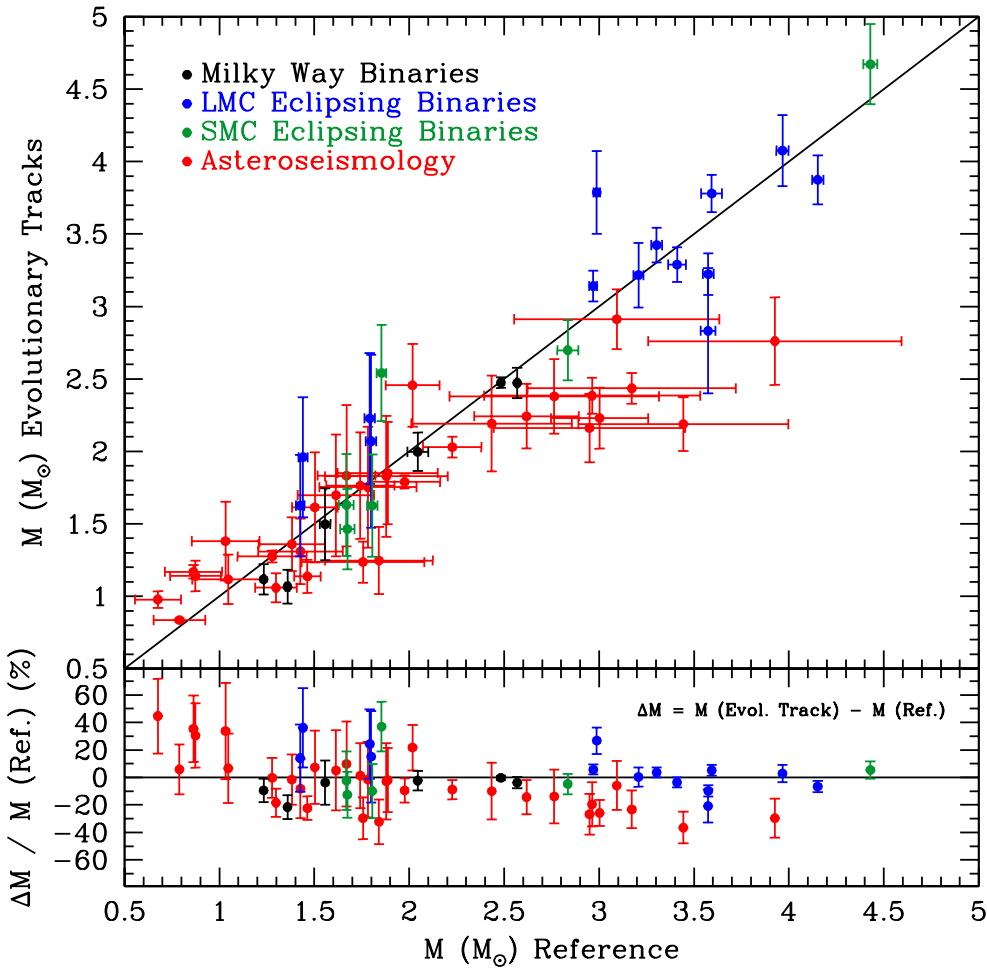}
\caption{Comparison between evolutionary track and reference masses. Milky Way binaries,
LMC EBs, SMC EBs and field stars with asteroseismology are shown by black, blue, green
and red points, respectively. Black solid lines represent a perfect agreement. The upper
panel shows a direct comparison of the masses. The lower panel shows the fractional difference between
the mass estimated from the evolutionary tracks and the reference mass. 
We can see there is a general good
agreement between the masses and no significant overestimate 
of the values by the evolutionary track method.}
\label{mass_comparison}
\end{figure*}

Our comparison between model-independent masses and the
corresponding values derived from evolutionary tracks (see Section \ref{tracks}) is
shown in Figure \ref{mass_comparison}. We can see there is an overall good
agreement and no systematic offsets for the entire mass range ($\sim$0.7 -- 4.5 \msun), 
even though our results are based on a  heterogeneous data set. 
The average absolute and percentage differences between evolutionary track and 
reference masses for the entire sample and the two subsamples (binaries and asteroseismic
targets) are shown in Table \ref{statistics}. We can see that the global average
differences are consistent with zero within the uncertainties, with 40 and 33 stars (i.e., 68\% and 56\% of the
sample, respectively) showing agreements between the two mass estimates within
20\% and 10\%, respectively. The separate results for the binary and asteroseismic samples are somewhat different. While
the average mass difference for the former is zero within the errors, it reveals that the evolutionary
track results are slightly underestimated for the latter. 
We therefore see no evidence that the models overpredict the masses of individual stars, 
in contrast to the concerns raised by \cite{lloyd13} and \cite{sw13}. Note, however, that the stars
analyzed here and in those works are in different evolutionary stages.

We performed weighted linear fits to the fractional residuals of the whole sample as 
well as the asteroseismic and binary subsamples individually. The weights were calculated as $1/\sigma_{total}^{2}$, where
$\sigma_{total}$ is the fractional uncertainty obtained with the propagation of errors for the difference of the two mass estimates. The results of the fits are listed in Table~\ref{statistics} and demonstrate that a slight trend is observed for the asteroseismic subsample, with the model grids underpredicting stellar masses by an amount that increases with mass. This is also clear from the lower panel in Figure \ref{mass_comparison}.

The results for the asteroseismic targets deserve some further discussion. Thirteen out of 33 stars have fractional differences larger than 20\%: Arcturus, 
$\beta$\,Aql, $\delta$ Eri, $\eta$ Ser, HD 170008, HD 170231, HD 178484, KIC 4044238, KIC 6101376, KIC 7909976, KIC\,8508931, KIC\,8813946, and $\xi$ Hya. 
\cite{kallinger10} derive a mass $M_\star = 0.80 \pm 0.20$~\msun\, for Arcturus (using \Dnu\, and an interferometric radius), which agrees better with the
value derived here from the evolutionary tracks ($M_\star = 0.98 \pm 0.06$~\msun\,). The adoption of the larger asteroseismic mass would reduce the fractional
difference from 45\% to 22\%. \cite{bruntt10} provide alternative masses for $\beta$~Aql, $\delta$ Eri and $\eta$ Ser: $M_\star = 1.26 \pm 0.18$~\msun\, 
(calculated from \numax, \teff\, and interferometric radius), $M_\star = 1.33 \pm 0.07$~\msun\, (calculated from \Dnu\, and interferometric radius) 
and $M_\star = 1.45 \pm 0.21$~\msun\, (calculated from \numax, \teff\, and luminosity), respectively. All these literature masses are in better agreement
with the ones obtained from evolutionary tracks and their usage would decrease the fractional differences from 31\%, 36\% and 30\% to 10\%, 12\% and 15\%,
respectively.

KIC\,8508931, has an alternative determination of the global asteroseismic parameters. Adopting the results from \cite{hekker11} (but using the same \teff\, as before and assuming uncertainties of 5\% in \numax\, and 2\% in \Dnu), we obtain a mass $M_\star = 2.54 \pm 0.43$~\msun, which reduces the discrepancy from 26\% to 12\%. For KIC\,8813946, \cite{huber12} found a conflict between the radii calculated for this star using different constraints (asteroseismic parameters and angular diameter from interferometry coupled with the parallax) and argue that this discrepancy could be indicative of a problem with the revised \textit{Hipparcos} parallax \citep{vanleeuwen07}. This problem would, in turn, affect our mass determined from the evolutionary tracks. Finally, the star $\xi$ Hya also has an alternative mass $M_\star = 2.89 \pm 0.23$~\msun\, from \cite{bruntt10}, calculated from \numax, \teff\, and its interferometric radius. The adoption of this mass would improve the agreement, with a decrease in the fractional difference from 23\% to 16\%. Unfortunately, we were not able to find alternative parameters for the stars HD 170008, HD 170231, HD 178484, KIC 4044238, KIC 6101376, and KIC 7909976.

If we replace the masses of the problematic stars with the alternative values mentioned 
above for six of them, the average differences become: $\langle\Delta M\rangle$ = -0.22 $\pm$ 0.07 \msun\ and 
$\langle\Delta M$/$M_{Ref.}\rangle$ = -7.18 $\pm$ 2.80\%. We can see that the fractional average difference 
became more negative because the large positive values of Arcturus, $\beta$\,Aql and $\delta$ Eri were decreased in the first case 
and replaced by negative offsets for the other two stars. The absolute average difference did not change significantly and no 
improvement in the standard deviations of the mean was observed. The parameters of the linear fit, on the other hand,
were slightly improved: $A = -7.98 \pm 2.34$, $B = 2.76 \pm 5.07$, $R^{2} = 0.274$ and $\sigma = 0.77$. Therefore, it is clear 
that part of the trend observed for the asteroseismic targets was caused affected by these six stars
with large discrepancies and, with the exception of $\eta$ Ser, located at the edges of the mass interval covered by the sample. 
The remaining trend
in the residuals can be mostly attributed to the other seven stars with large deviations and no alternative parameters to be tested.

Corrections to the scaling relations have been proposed by \cite{white11} and \cite{mosser13}, but further
tests are still necessary to confirm their accuracy. As a test, we applied the latter to 
the 33 stars in our asteroseismic sample (the former is valid only down to \teff\, = 4700 K and
some stars in the sample are cooler than this limit) and observed a typical decrease in the
asteroseismic masses of $\sim$5\%, which increases the average fractional difference 
$\langle\Delta M$/$M_{Ref.}\rangle$ from -4.29\% to 0.47\%. Although the agreement is better,
we decided to keep our original masses because there is still no consensus in the literature 
regarding the validity of the proposed corrections 
to the scaling relations or which one works best (e.g., \citealt{brogaard14}).

The exercise of testing alternative parameters could not be performed for the six binaries that have percentage differences 
larger than 20\%: RT CrB B, OGLE SMC 130.5 4296 A, OGLE LMC-ECL-01866 B, OGLE LMC-ECL-03160 A, 
OGLE LMC-ECL-09660 B and OGLE LMC ECL15620 A. To the best of our knowledge, there are
no alternative reference or input parameters that we could use to check if a better
agreement is obtained. Also, we were not able to observe any clear pattern in these
stars with larger deviations. They are all from different binary systems and
their companions seem to yield at least reasonable results (percentage differences
lower than 15\%). Moreover, they are not all the primary or secondary components in 
their systems. Thus, it is not feasible to attribute any problems to the analysis 
or the parameter determination for specific systems. The only detail worth noting is
that four of the six stars are from the LMC, i.e., are extragalactic in origin.

We also investigated if the larger fractional differences could be caused by the degeneracy between 
the two possible evolutionary stages for most stars in our sample: red giant branch (RGB) or the red clump (RC).
This degeneracy arises from the identical absolute magnitudes (luminosities) along the ascent of the RGB and stars on the 
horizontal branch, or red clump region of the H--R diagram. Uncertainties in the effective temperatures and metallicities of these
high-luminosity giants therefore results in a double-peaked
probability distribution function (PDF) (e.g., \citealt{dasilva06})
and this pattern can be indeed observed for many of our stars. A possible consequence of this issue
could be the choice of an incorrect mass associated with just one of the peaks 
(for instance, the mass corresponding to the maximum of the PDF),
creating a systematic offset for at least some of the stars in the sample. 

Our results do not show such offsets: evolutionary tracks are able to correctly recover the reference masses
with an average accuracy better than 10\% and no significant systematic trends. This good agreement is 
related to the way PARAM determines the best solution: it calculates mean values for the parameters from their PDFs, 
instead of choosing the most likely values, for example. Although this increases the uncertainties on the results,
it avoids attributing the parameters to a star given a choice of a specific evolutionary state. Therefore, the possible 
degeneracies are taken into account in our analysis and do not to seem to be the reason for the most discrepant cases.
We should also note that the differences between the two peaks of most of the double-peaked PDFs are within 68.3\%
confidence interval of the mean value. 

\begin{figure*}
%\epsscale{0.7}
\plotone{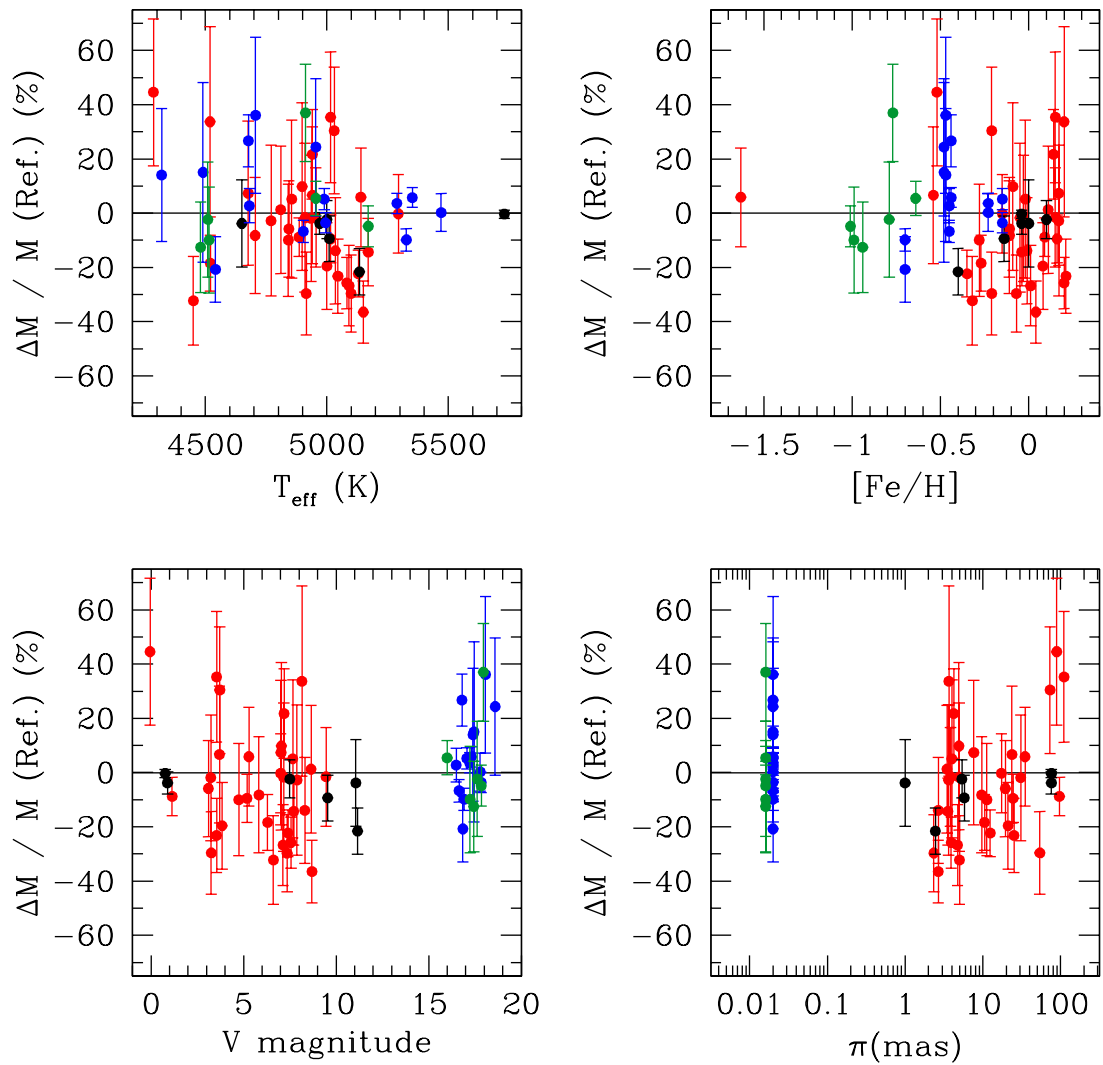}
\caption{Percentage difference between evolutionary track and reference masses as a
function of \teff\, (upper left panel), [Fe/H] (upper right panel), $V$ magnitude 
(lower left panel) and $\pi$ (lower right panel). Symbols and lines are
the same as in Figure \ref{mass_comparison}. We observe no systematic
trends for any of the parameters.}
\label{delta_masses}
\end{figure*}

As a final check on our results, we checked if the mass differences were a function of any of the 
parameters that were used as input to the evolutionary tracks method (\teff, [Fe/H],
$V$ magnitude and parallax). As can be seen in Figure \ref{delta_masses}, no systematic
trends are found for any of the parameters. The higher correlation coefficients $R^{2}$ 
of the weighted linear fits are 0.020 for the entire sample (in \teff), 0.014 for
the binaries (in [Fe/H]) and 0.124 (in parallax) for the asteroseismic targets.

\subsection{Radii}
\label{discussion_radii}

\begin{figure}
%\epsscale{0.7}
\plotone{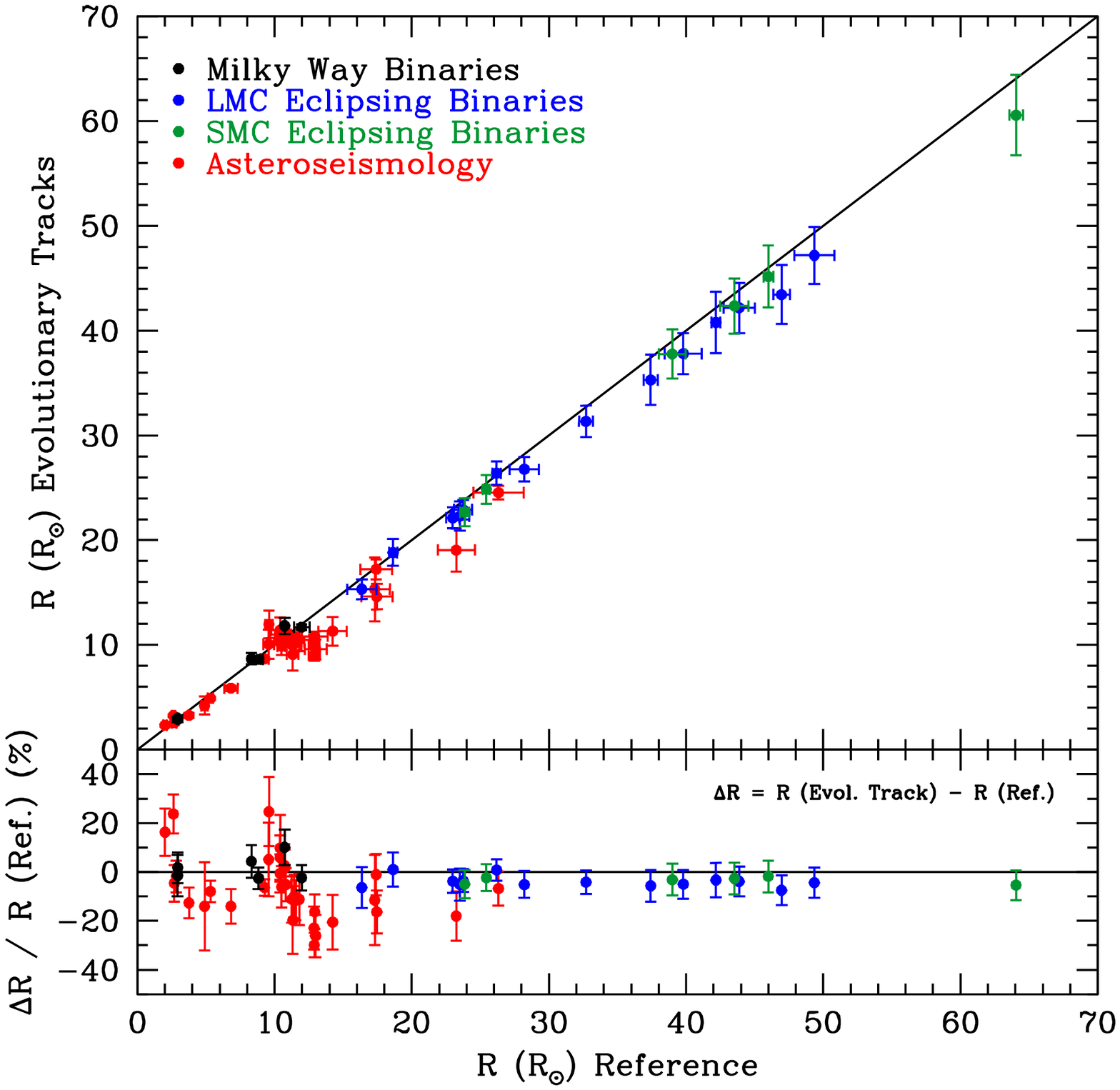}
\caption{Comparison between evolutionary track and reference radii. Symbols, lines and
panels are the same as in Figure \ref{mass_comparison}. We can see there is a general good
agreement between the radii with small offsets between the two sets of measurements.}
\label{radius_comparison}
\end{figure}

Our analysis of stellar radii closely follows that of the previous section. The comparison between reference radii and the
corresponding values estimated from evolutionary tracks (see Section \ref{tracks})
is depicted in Figure~\ref{radius_comparison}. Although the overall agreement is
good, we can see a global small offset ($\sim$5\%, see Table \ref{statistics}), 
with the evolutionary tracks underestimating the radii of most stars. 
There is also a clear difference between the samples of binaries and asteroseismic stars.
While the latter have a larger and mostly constant offset with a significant dispersion around it, the former
have a smaller dispersion around a small offset that slightly increases with increasing radius.  
It is worth noting though that these offsets are well within the uncertainties obtained 
in the radii from evolutionary tracks, as is also clear from Figure \ref{radius_comparison}.

Besides these interesting features, once again we can find asteroseismic targets with
percentage differences in radius larger than 20\%: $\beta$\,Aql, HD 169730, HD 170174, HD 170231, KIC\,8508931 and 
KIC\,8813946. Four of them also had problems in their masses. For $\beta$ Aql, our 
evolutionary track radius (3.25 $\pm$ 0.13 \rsun) is in very good agreement with the
interferometric value presented by \cite{bruntt10} (3.21 $\pm$ 0.13 \rsun). 
\cite{huber12} provide an interferometric radius for KIC\,8813946
(R = 12.0 $\pm$ 1.2 \rsun) which agrees very well with the value derived from
the evolutionary tracks (12.0 $\pm$ 1.3 \rsun). This could suggest a problem
in the asteroseismic global parameters for this star, but we should recall that the 
parallax is used as an input for both the interferometric and evolutionary track radii.
Thus, we can not discard a possible issue on the parallax (as suggested by 
\citealt{huber12}), which could be leading to two agreeing, but incorrect radii.
Finally, adopting for KIC\,8508931 the alternative global asteroseismic parameters from  
\cite{hekker11} (but using the same \teff\, as before and arbitrary uncertainties of 
5\% in \numax\, and 2\% in \Dnu) has a small impact on the radius, reducing the
percentage difference from 23\% to 19\%. Unfortunately, we were not able to find alternative 
parameters for the stars HD 169730, HD 170174 and HD 170231.

Corrections to the scaling relations from \cite{mosser13} were again applied as a test to 
asteroseismic sample. We observed a typical decrease in the asteroseismic radii of $\sim$2.5\%, 
which increases the average fractional difference $\langle\Delta R$/$R_{Ref.}\rangle$ for the asteroseismic targets 
from -6.58\% to -4.31\%. and improves the agreement between the two sets of radii. Despite this improvement, we decided to
keep our original radii for the same reason as explained for the masses. 

As in Section \ref{discussion_masses}, we investigated if the offsets between the two 
sets of radii were correlated with \teff, [Fe/H], $V$ magnitude and parallax. No systematic
behaviors were found in any case, with maximum correlation coefficients $R^{2}$ 
for the weighted linear fits equal to 0.044 for the entire sample (in \teff), 
0.204 for the binaries (in [Fe/H]) and 0.133 (in parallax) for the asteroseismic targets.

\subsection{Surface Gravities}
\label{discussion_gravities}

\begin{figure}
%\epsscale{0.7}
\plotone{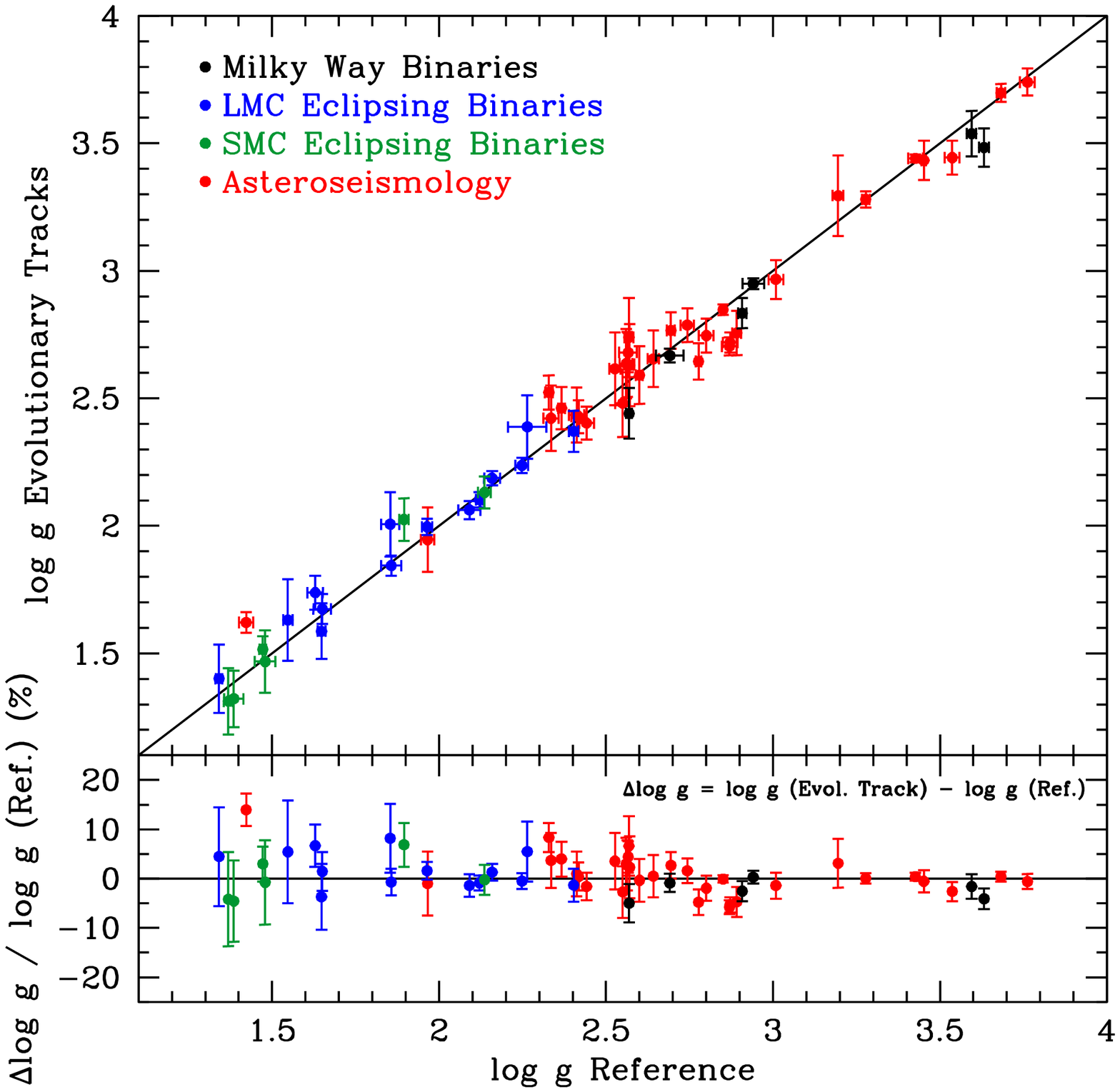}
\caption{Comparison between evolutionary track and reference surface gravities. Symbols,
lines and panels are the same as in Figure \ref{mass_comparison}. 
We can see there is a general good agreement between the gravities and no significant 
offsets introduced by by the evolutionary track method.}
\label{logg_comparison}
\end{figure}

The last of the three parameters that was derived from both model-independent and
-dependent methods is the surface gravity. The comparison between the two sets of
results is shown in Figure \ref{logg_comparison}. The agreement is remarkable, with
average absolute and percentage differences 0.01 $\pm$ 0.01 dex and 0.71 $\pm$ 0.51\%.
Such a good agreement is also observed for the subsamples of binaries and asteroseismic targets.
The more robust results for the surface gravity are expected since this parameter 
(rather than the mass or radius) is directly connected to the asteroseismic observables 
through the scaling relations (e.g., \citealt{kb95}).
Moreover, as noted by \cite{gai11}, the errors on asteroseismic masses and radii have a strong 
positive correlation that compensate each other to produce small uncertainties in $\log$ g.
For this reason, the usage of masses and radii derived from corrected scaling relations 
(following the prescription from \citealt{mosser13}) to calculate corrected surface
gravities leads to virtually equal values of $\log$ g.

As was the case for both mass and radius, no systematic trends were observed in the percentage differences
between the two sets of surface gravities as a function of any of the input parameters
(\teff, [Fe/H], $V$ magnitude and parallax). The higher correlation coefficients 
$R^{2}$ for the weighted linear fits were 0.039 for the entire sample (in [Fe/H]), 
0.041 for the binaries (in [Fe/H]) and 0.107 (in \teff) for the asteroseismic targets.

\subsection{Ages}
\label{discussion_ages}

The ages for all our benchmark stars were also given as an output by PARAM (see Table \ref{table_results_param}), 
but we were not able to compare them with the literature values because these were also determined using theoretical
isochrones. Although not all ages were helpful in assessing the performance of the evolutionary track
method, the ones derived for binary systems which have both members in our sample can provide
important information about the consistency of our results. The determination of ages for a binary
system usually requires that both stars are described by a single isochrone at a given age, which is a
reasonable assumption considering the stars were presumably born at the same time from the same molecular cloud. However, we did not impose
this constraint here and analyzed members of a given system as if they were individual isolated stars.
Thus, a good agreement between the ages derived for the stars in a binary system could be regarded
as a consistency check for the results obtained with the evolutionary track method.

\begin{figure}
%\epsscale{0.7}
\plotone{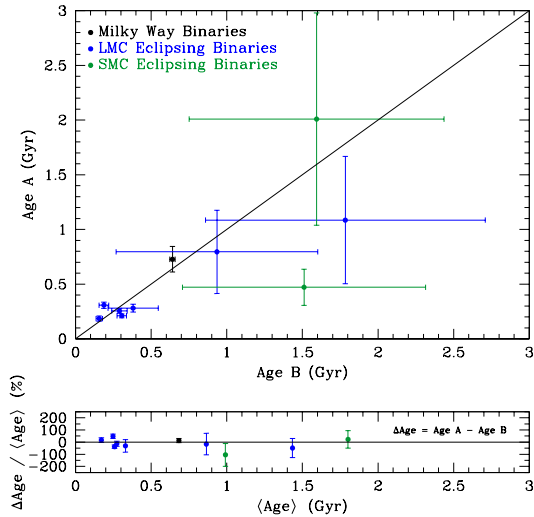}
\caption{Comparison between the ages of the members in binary systems. Symbols and lines
are the same as in Figure \ref{mass_comparison}. The upper panel shows a direct comparison 
of the ages. The lower panel shows the differences (in the sense A-B, relative to the mean age
of the system and as a percentage) between the two sets of ages relative to the average age of 
the system. We can see there is a general good agreement between the ages.}
\label{age_comparison}
\end{figure}

Our sample contains 10 systems for which both stars were analyzed: 7 from the LMC, 2 from the SMC and
one from the MW. The comparison between the ages of the members is shown in Figure \ref{age_comparison}.
We can see the overall agreement is good, with only three systems not having ages that agree within 
1$\sigma$: OGLE SMC 130.5 4296, OGLE LMC-ECL-09660 and OGLE LMC-ECL-26122. For OGLE SMC 130.5 4296, the 
A component has a model-dependent mass 37\% higher than the dynamical one, which is consistent with its 
lower age relative to the B component. A similar situation is observed for OGLE LMC-ECL-09660, with the 
only difference that it is the B component that has a larger model-dependent mass (by 27\%). 
For the system OGLE LMC-ECL-26122, both components present a good agreement between the two sets of mass 
(5\% for A and -4\% for B). However, the small offsets in opposite directions are enough to cause the age 
discrepancy, especially because of the relatively small errors. 

The average absolute and fractional differences between the ages of the components (in the sense A-B)
are -0.15 $\pm$ 0.13 Gyr and -15.08 $\pm$ 13.80 \% (where the errors are standard deviations of the mean).
A weighted linear fit to the residuals does not show any systematic trends ($R^{2}$ = 0.033). Thus, 
the consistency check with the ages provides additional support to the good performance of the evolutionary
track method for evolved stars.

\section{Conclusions}
\label{conclusion}

The determination of masses for evolved stars (subgiants and giants) is still a matter of
debate due to recent claims in the literature that the interpolation of observed properties 
(such as \teff, $\log(L/L_{\odot}$) and [Fe/H]) in a grid of evolutionary tracks could yield
systematically overestimated results. The accuracy of this method has not been extensively 
tested in this region of the Hertzsprung--Russell diagram.
In order to address this issue, we compiled a sample of 59 benchmark stars from the literature,
which includes 26 members of binary systems (in the MW, SMC and LMC) and 33 objects
with global asteroseismic parameters.

We determined model-dependent masses for all stars using the PARAM model-grid interpolation code and then compared with model-independent values coming from dynamical (binaries) or 
asteroseismic (single stars) mass measurements. We observed a very good agreement for the entire 
mass interval between $\sim$0.7 -- 4.5 \msun, even though a heterogeneous data set was used
in the study. The average absolute and fractional differences between model-dependent and reference masses
are $-0.10 \pm 0.05$~\msun\ and $-1.30 \pm 2.42$\%, respectively (where the errors are standard 
deviations of the means). No significant trends
in the residuals were found as a function of reference mass, \teff, [Fe/H], $V$ magnitude and
$\pi$ ($R^{2} \leq$ 0.12). Similar good agreements were also found for the radii and 
surface gravities. The analysis of the ages of binary systems provided an additional confirmation
that the results obtained with the evolutionary tracks were consistent.

The global results of our study suggest that the interpolation of observed parameters in a 
grid of evolutionary tracks, in particular the PARSEC set, is capable of providing accurate
and relatively precise masses, radii and surface gravities for evolved stars with different
effective temperatures and metallicities. We acknowledge, however, that much work remains to 
be done both observationally and theoretically in order to derive more accurate input 
parameters, improve the models and better understand why some specific 
stars present a poor agreement in our comparisons.  

With this in mind, our team is currently working on a follow-up study that will repeat the
analysis done here using different sets evolutionary tracks. This test could potentially determine the physics or free parameters that best reproduce the model-independent masses.
We are simultaneously working on the detailed characterization of other potential benchmark stars using
interferometry and asteroseismology, in a similar fashion as was done for HD\,185351
\citep{johnson14}. Finally, in the light of these new results, we plan to revisit the sample
of retired A stars from \cite{johnson10a}, presenting new analyzes of their spectroscopic and kinematic properties.

\acknowledgments

This research has made use of DEBCat and the SIMBAD database, operated at CDS, Strasbourg, France. 

The authors would like to thank Leo Girardi for providing the PARAM code as well as 
Daniel Huber, Victor Silva Aguirre, Thierry Morel and Nad\`ege Lagarde for kindly sharing their
asteroseismic data. We are also grateful to Leo Girardi, Willie Torres, Daniel Huber, Ben Montet,  
Enrico Corsaro, Saskia Hekker, Scott Fleming, Marc Pinnsoneault, Sarbani Basu 
and Andrea Miglio for helpful suggestions and discussions.

LG would like to thank the financial support from Coordena\c c\~ao de 
Aperfei\c coamento de Pessoal de N\'ivel Superior (CAPES), Ci\^encia sem Fronteiras,
Harvard College Observatory, and Funda\c c\~ao Lemann. JAJ is grateful for the
generous grant support provided by the Alfred P. Sloan and David \& Lucile 
Packard foundations.

\clearpage

\begin{turnpage}

\begin{deluxetable}{lrcrcrcr}
\tablecolumns{8}
%\rotate
\tabletypesize{\scriptsize}
\tablewidth{0pt}
\tablecaption{Properties of the Benchmark Stars.
\label{table_sample}}
\tablehead{
\colhead{Star} & \colhead{$V$} & \colhead{$E(B-V)$} &
\colhead{$\pi$} & \colhead{\teff} & \colhead{[Fe/H]} & 
\colhead{\numax} & \colhead{\Dnu} \\
\colhead{} & \colhead{} & \colhead{} & 
\colhead{(mas)} & \colhead{(K)} & \colhead{} & 
\colhead{($\mu$Hz)} & \colhead{($\mu$Hz)}
}
\startdata
\cutinhead{Binary Stars}
AI Phe B               &  9.542 $\pm$ 0.025 & 0.020 $\pm$ 0.020 &   5.78000 $\pm$ 0.37000 & 5010 $\pm$ 120 & -0.14 $\pm$ 0.10 & \nodata & \nodata \\
CapellaA               &  0.892 $\pm$ 0.016 & 0.000 &  75.99400 $\pm$ 0.08900 & 4970 $\pm$  50 & -0.04 $\pm$ 0.06 & \nodata & \nodata \\
CapellaB               &  0.763 $\pm$ 0.015 & 0.000 &  75.99400 $\pm$ 0.08900 & 5730 $\pm$  60 & -0.04 $\pm$ 0.06 & \nodata & \nodata \\
\cutinhead{Stars with Asteroseismology}
11 Com                 &  4.740 $\pm$ 0.020 & 0.016 $\pm$ 0.020 &  11.25000 $\pm$ 0.22000 & 4841 $\pm$ 100 & -0.28 $\pm$ 0.10 &  26.70 $\pm$  1.34 &  2.88 $\pm$ 0.06 \\
Arcturus               & -0.050 $\pm$ 0.010 & 0.000 &  88.83000 $\pm$ 0.54000 & 4286 $\pm$  30 & -0.52 $\pm$ 0.04 &   3.47 $\pm$  0.17 &  0.82 $\pm$ 0.02 \\
$\beta$ Aql               &  3.710 $\pm$ 0.009 & 0.000 &  73.00000 $\pm$ 0.20000 & 5030 $\pm$  80 & -0.21 $\pm$ 0.07 & 416.00 $\pm$ 20.80 & 29.56 $\pm$ 0.10 \\
\enddata
\tablecomments{A portion of the table is shown here for guidance regarding its form and content. The complete table is available at \url{https://drive.google.com/file/d/0B_C74xx43AOHTXBKNDh1YVI4RDA/view?usp=sharing}. \\
AI Phe B: Individual $V$ magnitude was calculated from the value of $M_{v}$ in \cite{andersen88} considering the reddening. $E(B-V)$ from \cite{hm84}. Parallax from \cite{torres10}. \teff\, and [Fe/H] from \cite{andersen88}. \\
Capella A and B: Individual $V$ magnitudes, parallaxes, \teff s and [Fe/H] from \cite{torres15}. $E(B-V)$ = 0.000 because d $\lesssim$ 60 pc. \\
%KIC 8410637: Individual $V$ magnitude, $E(B-V)$ and parallax from \cite{frandsen13}. \teff and [Fe/H] from \cite{hekker10}. \numax\, and \Dnu\, are the average values from the three determinations presented by \cite{hekker10}. \\
%RT CrB B: All parameters from \cite{sl03}. Individual $V$ magnitude was calculated from the value of $M_{v}$ in the paper. \\
%TZ For B: Individual $V$ magnitude was calculated from the value of $M_{v}$ in \cite{andersen91} considering the reddening. E(B-V) was obtained from $E(b-y) \approx$ 0.025 and the extinction ratio $k$ = 0.74 from \cite{rm05}. An arbitrary error of 0.02 was adopted. Parallax from \cite{torres10}. \teff\, and [Fe/H] from \cite{andersen91}. \\
%SMC stars: All parameters from \cite{graczyk14}. \\
%LMC stars: All parameters from \cite{pietrzynski13}. Arbitrary error $\sigma(V)$ = 0.02 was adopted for all stars. \\
11 Com: $V$ magnitude from SIMBAD. $E(B-V)$ converted from $A_{V}$ in \cite{takeda08} and an arbitrary error of 0.02 was adopted. Parallax from \cite{vanleeuwen07}. \teff\, and [Fe/H] from \cite{takeda08}. \numax\, and \Dnu\, from \cite{ando10}. Arbitrary errors of 5\% and 2\% in \numax\, and \Dnu, respectively. \\
Arcturus: $V$ magnitude from SIMBAD with an arbitrary error $\sigma(V)$ = 0.01. $E(B-V)$ = 0.000 because d $\lesssim$ 60 pc. Parallax from \cite{vanleeuwen07}. \teff\, and [Fe/H] from \cite{rap11}. \numax\, and \Dnu\, from \cite{lagarde15}. \\
$\beta$ Aql: $V$ magnitude from SIMBAD. $E(B-V)$ = 0.000 because d $\lesssim$ 60 pc. Parallax from \cite{vanleeuwen07}. \teff\, and [Fe/H] from \cite{bruntt10}. \numax\, and \Dnu\, from \cite{corsaro12}. Arbitrary error of 5\% in \numax. \\
}
\end{deluxetable}

%%%%%%%%%%

\clearpage

\begin{deluxetable}{lcccccccl}
\tablecolumns{9}
%\rotate
\tabletypesize{\scriptsize}
\tablewidth{0pt}
\tablecaption{Evolutionary parameters for the benchmark stars.
\label{table_results_param}}
\tablehead{
\colhead{} & \multicolumn{3}{c}{Reference Parameters} & \multicolumn{4}{c}{PARAM Results} & \colhead{} \\
\colhead{Star} & \colhead{$M$} & \colhead{$R$} & \colhead{$\log g$} & 
\colhead{$M$} & \colhead{$R$} & \colhead{$\log g$} & \colhead{Age} & \colhead{References} \\
\colhead{} & \colhead{(\msun)} & \colhead{(\rsun)} & \colhead{} & 
\colhead{(\msun)} & \colhead{(\rsun)} & \colhead{} & \colhead{(Gyr)} & \colhead{}
}
\startdata
\cutinhead{Binary Stars}
AI Phe B                & 1.234 $\pm$ 0.004 &  2.932 $\pm$  0.048 & 3.595 $\pm$ 0.014 & 1.118 $\pm$ 0.104 &  2.889 $\pm$  0.246 & 3.538 $\pm$ 0.089 &  6.616 $\pm$  2.122 & \cite{torres10} \\
Capella A              & 2.569 $\pm$ 0.007 & 11.980 $\pm$  0.570 & 2.691 $\pm$ 0.041 & 2.472 $\pm$ 0.104 & 11.690 $\pm$  0.285 & 2.668 $\pm$ 0.027 &  0.727 $\pm$  0.117 & \cite{torres15} \\
Capella B              & 2.483 $\pm$ 0.007 &  8.830 $\pm$  0.330 & 2.941 $\pm$ 0.032 & 2.474 $\pm$ 0.036 &  8.604 $\pm$  0.212 & 2.950 $\pm$ 0.021 &  0.641 $\pm$  0.017 & \cite{torres15} \\
\cutinhead{Stars with Asteroseismology}
11 Com                 & 2.435 $\pm$ 0.423 & 17.457 $\pm$  1.141 & 2.335 $\pm$ 0.022 & 2.192 $\pm$ 0.330 & 14.607 $\pm$  1.212 & 2.422 $\pm$ 0.128 &  0.931 $\pm$  0.381 & This work \\
Arcturus              & 0.676 $\pm$ 0.121 & 26.319 $\pm$  1.829 & 1.422 $\pm$ 0.022 & 0.978 $\pm$ 0.057 & 24.533 $\pm$  0.663 & 1.621 $\pm$ 0.040 &  8.616 $\pm$  1.653 & This work \\
$\beta$ Aql               & 0.872 $\pm$ 0.133 &  2.624 $\pm$  0.134 & 3.536 $\pm$ 0.022 & 1.140 $\pm$ 0.105 &  3.248 $\pm$  0.125 & 3.444 $\pm$ 0.066 &  5.859 $\pm$  1.868 & This work \\
\enddata
\tablecomments{A portion of the table is shown here for guidance regarding its form and content. The complete table is available at \url{https://drive.google.com/file/d/0B_C74xx43AOHTXBKNDh1YVI4RDA/view?usp=sharing}.}
\tablenotetext{a}{The reference parameters are from the analysis of the RVs and light curves of the eclipsing binary system. The parameters derived from asteroseismology are: M = 1.717 $\pm$ 0.142 \msun, R = 11.480 $\pm$ 0.434 \rsun\, and $\log$ g = 2.552 $\pm$ 0.006. The two sets of results agree within 1.6$\sigma$.}
\end{deluxetable}

\end{turnpage}

%%%%%%%%%%

\clearpage

\begin{deluxetable}{lll}
\tablecolumns{3}
%\rotate
\tabletypesize{\scriptsize}
\tablewidth{0pt}
\tablecaption{Stars not Included in the Analysis.
\label{table_removed_stars}}
\tablehead{
\colhead{Star} & \colhead{References} & \colhead{Note} \\
\colhead{} & \colhead{} & \colhead{} 
}
\startdata
ASAS 010538 B & \cite{ratajczak13} & [Fe/H] dependent on stellar evolution models \\
ASAS 182510 A,B & \cite{ratajczak13} & [Fe/H] dependent on stellar evolution models \\
ASAS 182525 A,B & \cite{ratajczak13} & [Fe/H] depends on stellar evolution models \\
$\beta$ Oph & \cite{kallinger10} & Error on reference mass $>$20\% \\
$\beta$ UMi & \cite{tarrant08} & Error on reference mass $>$20\% \\
CF Tau A & \cite{lacy12} & [Fe/H] dependent on stellar evolution models \\
$\epsilon$ Tau & \cite{ando10} & Suspicious (too high) reference mass \\ 
$\eta$ Her & \cite{ando10} & Suspicious (too high) reference mass \\ 
HD 50890 & \cite{lagarde15} & Error on reference mass $>$20\% \\
HD 169751 & \cite{lagarde15} & Error on reference mass $>$20\% \\
HD 170031 & \cite{lagarde15} & No independent parallax available \\
HD 170053 & \cite{lagarde15} & Error on reference mass $>$20\% \\
HD 175679 & \cite{lagarde15} & Error on reference mass $>$20\% \\
$\iota$ Dra & \cite{zechmeister08,baines11} & No value for \Dnu \\
Kepler-56 & \cite{huber13} & No independent parallax available \\
Kepler-91 & \cite{lb14} & No independent parallax available \\
KIC 3730953 & \cite{tt15} & Parallax error too large \\
KIC 5737655 & \cite{huber12} & Error on reference mass $>$20\% \\
KIC 6442183 & \cite{tian15} & No independent parallax available \\
KIC 9705687 & \cite{thygesen12} & Parallax error too large \\
KIC 11137075 & \cite{tian15} & No independent parallax available \\
KIC 11674677 & \cite{huber11} & Error on reference mass $>$20\% \\
M67 13 & \cite{kallinger10} & Distance is not model-independent \\
OGLE SMC113.3 4007 A & \cite{graczyk12} & Assumed [Fe/H] \\
OGLE LMC CEP0227 B & \cite{pilecki13,marconi13} & [Fe/H] depends on pulsation models \\
OGLE LMC-ECL-10567 A,B & \cite{pietrzynski13} & Remarks about the results for the system \\
\enddata
\tablecomments{There are 11 K giants in \cite{stello08}, but the values of \Dnu\, are not provided.}
\end{deluxetable}

%%%%%%%%%%

\clearpage

\begin{deluxetable}{lcrrrrrr}
\tablecolumns{8}
%\rotate
\tabletypesize{\scriptsize}
\tablewidth{0pt}
\tablecaption{Statistics for the comparison between evolutionary track and reference
parameters.
\label{statistics}}
\tablehead{
\colhead{Sample} & \colhead{N$_{Stars}$} & \colhead{$\langle\Delta Par\rangle$\tablenotemark{a}} &  
\colhead{$\langle\Delta Par$/$Par_{Ref.}\rangle$\tablenotemark{a}} & \colhead{A\tablenotemark{b}} &  
\colhead{B\tablenotemark{b}} & \colhead{$R^{2}$} & \colhead{$\sigma$} \\
\colhead{} & \colhead{} & \colhead{} & \colhead{(\%)} &
\colhead{} & \colhead{} & \colhead{} & \colhead{}
}
\startdata
\cutinhead{Par = M (\msun)}
All Stars        & 59 & $-$0.095 $\pm$ 0.052 & $-$1.30 $\pm$ 2.42 &     0.60 $\pm$ 1.65 &  $-$3.76 $\pm$ 4.79 & 0.002 & 1.32 \\
Binaries         & 26 &    0.041 $\pm$ 0.065 &    2.49 $\pm$ 2.99 &     0.57 $\pm$ 1.93 &  $-$2.24 $\pm$ 5.76 & 0.004 & 1.34 \\
Asteroseismology & 33 & $-$0.202 $\pm$ 0.073 & $-$4.29 $\pm$ 3.58 & $-$11.53 $\pm$ 2.93 &    12.12 $\pm$ 6.75 & 0.333 & 0.89 \\
\cutinhead{Par = R (\rsun)}
All Stars        & 59 & $-$1.002 $\pm$ 0.172 & $-$4.81 $\pm$ 1.32 &     0.00 $\pm$ 0.07 & $-$5.07 $\pm$ 1.86 & 0.000 & 1.26 \\
Binaries         & 26 & $-$1.033 $\pm$ 0.214 & $-$2.57 $\pm$ 0.73 &  $-$0.11 $\pm$ 0.04 &    0.45 $\pm$ 1.21 & 0.268 & 0.49 \\
Asteroseismology & 33 & $-$0.978 $\pm$ 0.261 & $-$6.58 $\pm$ 2.25 &  $-$0.74 $\pm$ 0.36 & $-$0.54 $\pm$ 4.10 & 0.122 & 1.44 \\
\cutinhead{Par = $\log$ g}
All Stars        & 59 &    0.010 $\pm$ 0.011 &    0.71 $\pm$ 0.51 &  $-$1.14 $\pm$ 0.60 &    3.04 $\pm$ 1.77 & 0.061 & 1.22 \\
Binaries         & 26 &    0.003 $\pm$ 0.015 &    0.49 $\pm$ 0.74 &  $-$2.05 $\pm$ 0.68 &    4.74 $\pm$ 1.70 & 0.277 & 0.75 \\
Asteroseismology & 33 &    0.015 $\pm$ 0.016 &    0.89 $\pm$ 0.70 &  $-$1.28 $\pm$ 1.14 &    3.70 $\pm$ 3.56 & 0.040 & 1.48 \\
\enddata
\tablenotetext{a}{$\Delta Par$ = $Par_{Trk.}$ - $Par_{Ref.}$, where Par can be mass, radius or surface gravity. The uncertainties are standard deviations of the means.}
\tablenotetext{b}{Coefficients of the linear fit $y = Ax + B$.}
\end{deluxetable}

\global\pdfpageattr\expandafter{\the\pdfpageattr/Rotate 90}

\end{document}